\documentclass[showpacs,twocolumn,floatfix,superscriptaddress]{revtex4}
\usepackage{graphicx}
\usepackage{amsmath}
\begin{document}
\title{Interaction effects on  counting statistics and the  transmission distribution}
\author{M.\ Kindermann}
\affiliation{Instituut-Lorentz, Universiteit Leiden, P.O. Box 9506, 2300 RA
Leiden, The Netherlands}
\author{Yu.\ V.\ Nazarov}
\affiliation{Department of Nanoscience, Delft University of Technology,
Lorentzweg 1, 2628 CJ Delft, The Netherlands}
\date{April 2003}
\begin{abstract}
We investigate the effect of weak interactions on the full counting statistics of
charge transfer through an arbitrary mesoscopic conductor. We show  that the main
effect can be incorporated into an  energy dependence of the transmission eigenvalues
and study this dependence in a non-perturbative approach. 
An unexpected result
is that all mesoscopic conductors behave at low energies like
either a single or a double tunnel junction, which divides them into
two broad classes.
\end{abstract}
\maketitle

It has been shown that at low energy scales the relevant 
part of the electron-electron interaction in mesoscopic conductors
comes from their  electromagnetic environment \cite{OldNazarovIngold}.
The resulting  dynamical Coulomb blockade has been thoroughly investigated for
tunnel junctions \cite{Dev90}. The measure of the interaction strength is the external 
impedance $Z(\omega)$ at the frequency scale $\Omega = {\rm max}(eV,k_B T)$ determined by either the voltage $V$ at the conductor or its temperature $T$.   If $z \equiv G_Q Z(\Omega)\ll 1$ (with the conductance quantum $G_Q=e^2/2\pi\hbar$)  the interaction is weak, otherwise Coulomb effects strongly suppress  electron
transport.    

A tunnel junction is the simplest mesoscopic conductor. 
An arbitrary mesoscopic conductor in the absence of interactions
is characterized by its scattering matrix or, most conveniently,
by a set of  transmission eigenvalues $T_n$ \cite{Bee98}.
This Landauer-B\"{u}ttiker approach to mesoscopic transport can be
extended to access the full counting statistics (FCS) of charge transfer
\cite{Lev93}. The FCS contains not only the average current but also 
 current noise and all higher moments of current correlations in a compact 
and elegant form. 

Interaction effects on 
general mesoscopic conductors are difficult to quantify
for arbitrary $z$. 
For $z \ll 1$, one can employ perturbation theory  
to  first order in $z$ \cite{OldNazarov2}. 
Recent work \cite{Zai01,Yey01}
associates the resulting  interaction correction to the conductance 
with  shot noise properties
of the conductor. The interaction correction to noise is associated
with the third cumulant of charge transfer \cite{Gal02}.
This motivates us to study the interaction correction to
all cumulants of charge transfer, i.e. 
to the  FCS. The recent experiment \cite{Cro02} 
addresses the correction to 
the conductance at arbitrary transmission.

A tunnel junction in the presence of an electromagnetic environment exhibits an
anomalous power-law I-V characteristic, $I(V) \simeq V^{2z+1}$.
The same power law behavior is typical for 
tunnel contacts between one-dimensional interacting electron systems, the so-called Luttinger liquids
\cite{Kan92}. It has  also been found  for contacts with arbitrary
transmission between single-channel conductors in the limit of  weak interactions
\cite{Mat93}. In this case, the interactions have been found to renormalize the
transmission.

In this Letter we study the effects of weak interactions $z\ll 1$ on the  FCS
of  a phase
coherent multi-channel conductor. In the energy range below the Thouless energy that we restrict our analysis to, its transmission probabilities $T_n$ are energy independent in the absence of interactions. We first analyze the interaction correction
to  first order in $z$. We identify an elastic and an inelastic 
contribution. The elastic contribution comes with
a logarithmic factor that diverges at low energies suggesting
that even  weak interactions can suppress  electron transport
at sufficiently low energies. To quantify this we sum up
interaction corrections to the FCS of all orders in $z$ by a
renormalization group analysis. We show that the result 
is best understood  as
 a renormalization of the transmission eigenvalues similar
to that proposed in \cite{Mat93}. The renormalization brings about
an energy dependence of the transmission eigenvalues according 
to the flow equation 
\begin{equation}
\label{main}
\frac{ d T_n(E)}{d {\rm \ln} E} = 2 z\,  T_n(E) [ 1 - T_n(E)]. 
\end{equation}
To calculate transport properties in the presence of interactions, 
one evaluates $T_n(E)$ at the energy $E \simeq \Omega$.

With  relation  (\ref{main})  
we explore the effect of interactions on 
the distributions of
transmission probabilities for various types of mesoscopic conductors. 
In general, their conductance $G$  and their  noise properties display a complicated
behavior at $z |\ln E| \simeq 1$ that depends on details
of the conductor. However, in the limit of very
low energies $z |\ln E| \gg 1$ we find only two possible scenarios.
The first one is that the conductor behaves like a {\it single} tunnel junction
with $G(V) \simeq V^{2z}$. In the other scenario, the transmission
distribution approaches that of a symmetric  {\it double}
 tunnel junction. 
The conductance scales then  as $G(V) \simeq V^{z}$.
Any  given conductor follows one of the two scenarios. This divides
all mesoscopic conductors into two broad classes.

We start out by evaluating the interaction correction to the FCS to first order in $z$.
We analyze  a simple circuit that
consists of a mesoscopic conductor in series with an external 
resistor $Z(\omega)$ biased with a  voltage source $V$ (Fig. 1). 
For this we employ a  non-equilibrium Keldysh action technique \cite{art6}.
\begin{figure}
 \label{fig1} 
\includegraphics[width=6cm]{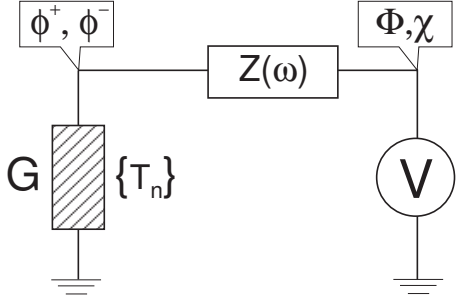}
\caption{Phase-coherent conductor (conductance $G$) in an electromagnetic 
environment  (impedance $Z(\omega)$).
We formulate the quantum dynamics of the system  in terms of
the fluctuating fields $\phi^{\pm}(t)$.} 
\end{figure}
Within this approach, one represents the generating function 
${\cal F}(\chi) $ of  current
fluctuations in the circuit as  a path integral  
over the fields $\phi^{\pm}(t)$ defined on the time interval $[0,\tau]$,
that represent the fluctuating voltage in the node 
shared by mesoscopic conductor
and external resistor.
 The path integral representation of   ${\cal F}(\chi)$ reads
\begin{eqnarray} 
\label{eq:path} 
{\cal F}(\chi)& =&  \int{
{\cal D} \phi^+ {\cal D} \phi^- \exp\left\{ -i
{\cal S}_{\rm c} \left[\phi^+,\phi^-\right]\right. } 
\nonumber \\ 
&& \left.-i{\cal S} _{\rm env}
\left[\Phi+\chi/2 -\phi^+ ,\Phi-\chi/2 -\phi^-\right]\right\} 
\end{eqnarray} 
where $d\Phi(t)/dt \equiv e V$. Derivatives of ${\cal F(\chi)}$ with respect to $\chi$
at $\chi=0$ give the moments 
of the charge transferred through the circuit  
during a time interval of length $\tau$.
The Keldysh action is  a sum of two terms ${\cal S}_{\rm env}$ and  ${\cal S}_{\rm c}$
describing the environment and the mesoscopic conductor respectively.

We assume a linear electromagnetic environment that
can be fully characterized by its impedance $Z(\omega)$ and its temperature $T$. Its  action 
is bilinear in $\phi^{\pm}$,
\begin{eqnarray} &&{\cal S} _{\rm env}= \frac{1}{2 \pi}
\int_0^{\tau} dt \int_0^{\tau}  dt'
\left[
\phi^+(t) A^{++}(t-t')  \phi^+(t')\right.\nonumber \\ 
&& \left.\;\;+\phi^+(t) A^{+-}(t-t') \phi^-(t')+
\phi^-(t) A^{--}(t-t')  \phi^-(t') \right] \nonumber
\end{eqnarray}
with
\begin{eqnarray}
A^{++}(\omega)&=& - i \omega[z^{-1}(\omega) + 2 N(\omega) {\rm Re}\,z^{-1}(\omega)] \nonumber\\
A^{+-}(\omega)&=& 4 i \omega N(\omega) {\rm Re}\,z^{-1}(\omega)  \nonumber\\
A^{--}(\omega) &=&-[ A^{++}(\omega)]^*. 
\end{eqnarray} 
Here, $N(\omega)\equiv \left\{ \exp[
\omega/k_B T]-1\right\}^{-1}$ is the  Bose-Einstein distribution function
and  $z(\omega)= G_Q Z(\omega)$  the  dimensionless frequency-dependent impedance.

The action ${\cal S}_{\rm{c}}$ of the mesoscopic conductor can be 
 expressed
in terms of  Keldysh Green functions 
${\check G}_{R,L}$ (the "check" denotes $2 \times 2$
matrices in Keldysh space) of electrons in the  two 
reservoirs adjacent to the conductor \cite{Naz99}. It takes the form of a trace over frequency and Keldysh indices,
\begin{eqnarray} \label{eq:action} {\cal
S}_{\rm c}  = \frac{i}{2}
\sum_n {\rm Tr} \;\ln \left[1 + \frac{T_n}{4} \left( \left\{ {\check G}_{\rm
L} ,{\check G}_{\rm R} \right\} -2 \right) \right]
\end{eqnarray} 
and depends on the set of transmission eigenvalues $T_n$ that characterizes the
conductor. The fields $\phi^{\pm}(t)$ enter the expression as a gauge transform
of $\check G$ in one of the reservoirs,  
\begin{eqnarray}
&&{\check G}_{\rm
R}=   {\check G}^{\rm res} \;\;\; {\rm and} \;\;\;\; {\check G}_{\rm
L}(t,t')= \nonumber \\
&& {\textstyle \left[ \begin{array}{cc}e^{ i\phi^{+}(t)} & 0 \cr
0 & e^{i\phi^{-}(t)} \end{array}\right] }
{\check G}^{\rm res}(t-t')
{\textstyle \left[ \begin{array}{cc} e^{-i\phi^{+}(t')} & 0 \cr
0 & e^{-i\phi^{-}(t')} \end{array}\right]} . \nonumber \\
\end{eqnarray}
 $G^{\rm res}$ is the equilibrium
 Keldysh Green function  
\begin{equation}
 \label{eq:xiG} 
{\check G}^{\rm res}(\epsilon)= 
\left(
\begin{array}{cc} 1 - 2 f(\epsilon) & 2 f(\epsilon) \\ 2[1 - f(\epsilon)] & 2
f(\epsilon)-1 \end{array} \right),
 \end{equation} 
$f(\epsilon)$ being the equilibrium electron distribution function.

This defines our model that is valid for any 
external impedance but is hardly tractable in the general case.
We proceed with perturbation theory in $z$ assuming
that $z \ll 1$. To zeroth order in $z$ the
fields $\phi^{\pm}(t)$ 
do not fluctuate and are fixed to $eVt \pm \chi/2$.
Substituting this into Eq. (\ref{eq:action}) we recover Levitov's
formula for non-interacting electrons,
\begin{eqnarray} 
\label{eq:zeroorder} &&  {\rm ln} \,{\cal F}^{(0)}(\chi)\equiv -i \tau{\cal S}^{(0)} 
\left(\chi\right) =  \tau \int{\frac{d\epsilon}{2 \pi} \sum_n \ln\left\{1
\right.} \nonumber \\ &&\left. + T_n\left[ (e^{i \chi}-1) f_{\rm
L}(1 - f_{\rm R}) + (e^{-i\chi}-1) f_{\rm R}(1 - f_{\rm L})\right]
\right\} \nonumber \\
 \end{eqnarray} 
($f_R \equiv f$ and $f_L(\epsilon)\equiv f(\epsilon - eV)$). Interaction effects manifest themselves at higher orders in $z$.   To assess the first order  correction,
we expand the non-linear ${\cal S}_{\rm c}$ to second order in the 
fluctuating fields $\phi^{\pm}(t)$. We integrate it over $\phi^{\pm}$
with the weight given by ${\cal S}_{\rm env}$. The expression
for the correction can be presented as \cite{z0}
\begin{eqnarray}
\label{eq:firstorder} 
&& \ln {\cal F}^{(1)}(\chi) = -i \tau \int_0^{\infty}{d\omega
\,\frac{{\rm Re}\, z(\omega)}{\omega} \left\{ [2N(\omega)+1] {\cal S}^{(1)}_{\rm el}(\chi) \right.}
\nonumber \\ 
\label{eq:1st1}  && \;\;\;\;\;\;\;\;\;\;\left.+ N(\omega)
 {\cal S}^{(1)}_{\rm in}(\omega,\chi) + [N(\omega)+1]
{\cal S}^{(1)}_{\rm in}(-\omega,\chi)\right\}. \nonumber \\
\end{eqnarray} 
The three terms in square brackets correspond to   {\it elastic} electron transfer,
inelastic  transfer with absorption of energy $\hbar \omega$ from
the environment, and  inelastic electron transfer with emission 
of this energy  respectively. It is crucial  to note that inelastic processes can only occur at frequencies $\omega \lesssim \Omega$ and that their contribution
to the integral is thus restricted to this frequency range. In contrast, elastic contributions come primarily from  frequencies exceeding
the scale $\Omega$. If $z={\rm const}(\omega)$ for $\omega \lesssim \Lambda$,
the elastic correction diverges logarithmically, its magnitude being 
$\simeq z \ln \Lambda /\Omega$. This suggests that
i. the elastic correction is more important than the inelastic one
and ii. a small value of $z$ can be compensated for by a large logarithm,
indicating the breakdown of perturbation theory.   
The upper cut-off energy $\Lambda$ is set either by the inverse $RC$-time of  
the environment circuit or 
the Thouless energy of the electrons in the mesoscopic conductor. 

The concrete expression for ${\cal S}^{(1)}_{\rm in}$ reads 
\begin{eqnarray}
\label{eq:1st2}  {\cal S}^{(1)}_{\rm in}(\omega,\chi)& =& i  \sum_n
\int{ \frac{d\varepsilon}{2\pi} \,\left\{ D_n D_n^+\left[ T_n (f_L-f_L^+)
 \right. \right. } \nonumber \\
&&\left. \left. + 2 T_n (e^{i\chi}-1)f_L(1-f^+_R)   \right. \right. \nonumber \\
&& \left. \left. + 2 T_n^2  (\cos \chi
-1)   f_L (1-f_L^+)(f_R^+-f_R)  \right]\right. \nonumber \\
&& \left. +  T_n D_n + (1-D_n)(1-D_n^+) \right\}  \nonumber \\
&& + \left\{ \begin{array}{cc} {\rm R} \leftrightarrow {\rm L} \\  {\chi \leftrightarrow -\chi} \end{array} \right\},
 \end{eqnarray}
where we have introduced the functions
 \begin{eqnarray} \label{eq:1st4} 
D_n&=& \left\{ 1+ T_n \left[ f_L(1-f_R)(e^{i \chi}-1) \right. \right.
\nonumber \\ && \;\;\;\;\;\;\;\;\;\;\; \left. \left. +
f_R(1-f_L)(e^{-i\chi}-1)\right]\right\}^{-1}
 \end{eqnarray} 
and the notation 
\begin{equation} 
f^{+}(\varepsilon)=f(\varepsilon+\omega), \;\;
D_n^+(\varepsilon)=D_n(\varepsilon+\omega). 
\end{equation}
 We do not
analyze ${\cal S}^{(1)}_{\rm in}$ further in this Letter and instead turn to the
analysis of the elastic correction.

It is important that the explicit 
form of the elastic correction can be presented
as 
\begin{eqnarray} \label{eq:zerotemp} 
{\cal S}_{\rm el} &=&  \sum_n \delta T_n 
\frac{\partial {\cal S}^{(0)} }{\partial T_n} \nonumber \\
{\rm with} \; \delta T_n &=& - 2 T_n (1-T_n). 
\end{eqnarray}
This suggests that the main effect of interactions is to change
the transmission coefficients $T_n$. It also suggests that 
we can go beyond perturbation theory  by 
a renormalization group analysis that involves the  $T_n$ only.
In such an analysis one concentrates at each renormalization step  on the
"fast" components of $\phi^{\pm}$ with frequencies in a narrow
interval $\delta \omega$ around the running cut-off frequency $E$.
Integrating out these fields one obtains a new action for the "slow" fields. 
Subsequently one reduces $E$ by $\delta\omega$
and repeats the procedure until the running cut-off approaches $\Omega$.
We find that at each step of renormalization the action indeed retains
the form given by Eq. (\ref{eq:action}) and only the  $T_n$ change,
provided $z \ll {\rm min}\{1,G_Q/G\}$. The resulting energy dependence of the $T_n$ obeys Eq. (\ref{main}). The strict proof of this 
involves manipulations on the  action (\ref{eq:action}) 
with time dependent arguments
$\phi^{\pm}(t)$. It is very technical and  we do not give it here. 
The approximations that we make in this renormalization procedure amount to a
summation of the leading logarithms in every order of  the  perturbation series.
   
In the rest of the Letter 
we analyze the consequences of Eq. (\ref{main}) for various mesoscopic
conductors.
Equation (\ref{main}) can be explicitly integrated  to obtain 
\begin{equation}
\label{eq:tofE}
 T_n(E)  = \frac{\xi T^{\Lambda}_n }
{1 -T^{\Lambda}_n \left(1-\xi\right)}, \;\; 
\xi \equiv \left( \frac{E}{\Lambda}\right)^{2z}
\end{equation}
in terms of the "high energy" (non-interacting) transmission eigenvalues $T^{\Lambda}$.
A mesoscopic conductor containing many transport channels  is most conveniently characterized by
the distribution $\rho_{\Lambda}(T)$ of its transmission eigenvalues \cite{general}.
It follows from Eq. (\ref{eq:tofE}) that the effective transmission distribution at the energy scale $E$ reads
\begin{equation}
\label{eq:rho}
\rho_{E}(T)= \frac{\xi}{[\xi+T(1-\xi)]^2} \rho_{\Lambda}\left( 
\frac{T}{\xi+T(1-\xi)}\right).
\end{equation}
We now analyze its low energy limit $\xi \rightarrow 0$. Any given
transmission eigenvalue will approach zero in this limit. Seemingly
this implies that for any conductor the transmission distribution
 approaches  that of a tunnel junction, 
so that all $T_n \ll 1$.
The overall conductance would be proportional to $\xi$ in accordance
with Ref. \cite{OldNazarov2}. 

Indeed, this is one of the possible scenarios. A remarkable exception
is the case that the non-interacting $\rho_{\Lambda}$ has an inverse square-root
singularity at $T \rightarrow 1$. Many mesoscopic conductors
display this feature, most importantly diffusive ones \cite{general}.
In this case, the low-energy  transmission
distribution approaches a limiting function
\begin{equation}
\label{eq:limiting}
\rho_*(T) \propto \sqrt{\frac{\xi} {T^3(1-T) }}.
\end{equation}
The conductance scales like $\xi^{1/2}$. $\rho_*$
is known to be the transmission distribution of a {\it double} tunnel
junction: two identical tunnel junctions in series \cite{Melsen}.
Indeed, one checks that for a double tunnel  junction the form
of the transmission distribution is unaffected by interactions.
This sets an alternative low-energy scenario. We are not aware of
transmission distributions that would give rise to other scenarios.

We believe that this is an important general result in the theory of
quantum transport and suggest now a qualitative explanation.
The statement is  that the conductance of a phase-coherent
conductor at low voltage and temperature $\Omega \ll \Lambda$
asymptotically obeys  a power law with an exponent that  generically takes two values, 
\begin{equation}
\label{eq:scaling}
G \propto \left(\frac{\Omega}{\Lambda}\right)^{2z} ,\;\; {\rm or} \;\;\;G \propto \left(\frac{\Omega}{\Lambda}\right)^{z}.
\end{equation}   
For tunneling electrons the exponent is $2z$. 
An electron traverses the conductor in a single leap.
The second  possible exponent  
$z$  has been  discussed in the  literature as well, 
in connection with  { \it resonant} tunneling through a {\it double} tunnel barrier in the presence of
interactions \cite{Kan92}. This resonant tunneling takes place via intermediate
discrete states contained between the two tunnel barriers. 
  The halved exponent $\alpha=z$ occurs in  the regime of the so-called {\it successive} 
electron tunneling. In this case, the  electron first jumps over
one of the barriers ending up in a discrete state. Only in a second jump over
the  second barrier the charge transfer is completed.
Since it takes two jumps to transfer a charge, 
the electron feels only half the counter voltage due to interactions with electrons in the environmental impedance $Z$ at each hop. Consequently, the exponent  at each jump takes  half the  value for direct tunneling.
Our results strongly suggest that this transport mechanism
is not restricted to resonant tunneling systems, or, in other words, that   resonant tunneling can occur in  systems of 
a more generic nature than generally believed. 
As far as transport is concerned, a mesoscopic conductor is characterized by its scattering
matrix regardless of the details of its inner structure.
 In this approach  it is not even obvious that 
the conductor can accommodate  discrete   states.
Nevertheless,  the transmission distribution of this scattering matrix does depend on the internal structure of the conductor.  The inverse square root singularity of this distribution at $T\to1$ for a double tunnel barrier is   due to   the formation of Fabry-Perot  resonances  between the two  barriers. Probably similar resonances are at the origin of the same singularity for  more complicated mesoscopic conductors with multiple scattering. They are then  the intermediate
discrete states that give rise to the modified scaling of the conductance in  presence of interactions. One may speculate that in diffusive conductors these resonances are the so-called ''nearly localized states'' found in  \cite{Khm95}. 

From equation (\ref{eq:tofE}) one concludes
that the resonant tunneling scaling holds only if $G(E) \gg G_Q$
so that many transport channels  contribute to the conductance.
At sufficiently small energies, $G(E)$ becomes of the order of $G_Q$.
All  transmission eigenvalues are then small and the conductance crosses over to  the  tunneling scaling.

\begin{figure} 
\includegraphics[width=8cm]{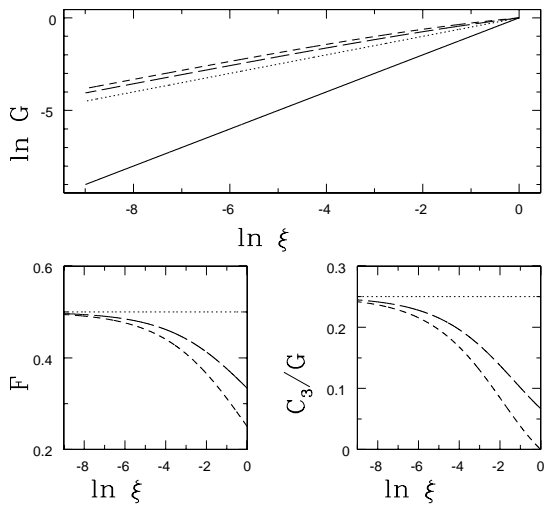}
\caption{Renormalization of  conductance $G$ (logarithmically), 
Fano factor $F=C_2/G$ and  third cumulant $C_3$ normalized by the
conductance in a tunnel junction (solid line), a double tunnel barrier
(dotted line), a double point contact (short dashed line), and a
diffusive conductor (long lashed line). }  \label{fig2}
\end{figure} 

With Eqs.\  (\ref{eq:tofE}) and  (\ref{eq:rho})
 we can  evaluate the transmissions and the  FCS in the intermediate regime $\xi \sim 1$.
 Fig. \ref{fig2}  shows the results for  
the first three cumulants of charge transfer $C_n$ for several types
of conductors whose non-interacting  transmission distributions
$\rho_{\Lambda}(T)$ are known. Apart from the tunnel contact all these conductors
approach the resonant tunneling scaling with noise properties of a double tunnel barrier  at very small bias voltage 
($\xi \ll 1$).

We remark, that Eq.\ (\ref{eq:zerotemp}) generalizes the statements made in \cite{Yey01,Gal02}: At zero temperature, the interaction correction to the n-th cumulant of transferred charge is proportional to the (n+1)-th cumulant. 

To conclude, we have investigated the effects of interactions on 
the FCS of a Landauer-B\"{u}ttiker conductor and found that their  main effect can be incorporated into  an energy dependence of the transmission eigenvalues.
For an arbitrary conductor, the conductance in the low-energy limit
obeys one of two  generic  scaling laws.  

The authors acknowledge useful discussions with C. W. J. Beenakker,  K. B. Efetov, D. Esteve,
L. I. Glazman, L. S. Levitov, and A. D. Zaikin. 
This work was supported by the Dutch Science Foundation NWO/FOM.


\begin{thebibliography}{99}
\bibitem{z0} 
We disregard corrections $\propto z(0)$ that  account for
the division of the voltage between  mesoscopic conductor and
external impedance (see \cite{art6} for details).
\end{thebibliography}
\end{document}